%% file: main.tex
\documentclass[aps,pra,twocolumn,amsmath, amssymb, superscriptaddress]{revtex4-2}
\usepackage{graphicx} 

\usepackage[dvipsnames]{xcolor}
\definecolor{myblue}{RGB}{39, 144, 219}
\definecolor{myred}{RGB}{219, 64, 61}

\usepackage{hyperref}
\hypersetup{
	colorlinks=true,
	linkcolor=myred,
	filecolor=magenta,
	urlcolor=myblue,citecolor=myblue}

\usepackage{orcidlink}

\usepackage[capitalize]{cleveref}

\newcommand{\VR}{r_v}

\begin{document}

\title{Emergent aggregation from collective foraging}
\author{Gorka Mu\~noz-Gil \orcidlink{0000-0001-9223-0660}}
\email{gorka.munoz-gil at uibk.ac.at}
\affiliation{University of Innsbruck, Department for Theoretical Physics, Technikerstr. 21a, A-6020 Innsbruck, Austria}

\author{Andrea Lopez-Incera \orcidlink{0000-0002-0522-6610}}
\affiliation{Universitat Autònoma de Barcelona, Departamento de Didáctica de la Matemática y las Ciencias Experimentales, 08193 Bellaterra (Cerdanyola del Vallès), Spain
}

\author{Vide Ramsten \orcidlink{0009-0005-6474-0846}}
\affiliation{Department of Physics, University of Gothenburg, Origovägen 6B, 41296, Göteborg, Sweden}

\author{Giovanni Volpe \orcidlink{0000-0001-5057-1846}}
\affiliation{Department of Physics, University of Gothenburg, Origovägen 6B, 41296, Göteborg, Sweden}

\author{Thomas Müller \orcidlink{0000-0003-1225-1483}}
\affiliation{Department of Philosophy, Box 17, University of Konstanz, 78457 Konstanz, Germany}

\author{Hans J.~Briegel \,\orcidlink{0000-0002-9065-1565}}
\affiliation{University of Innsbruck, Department for Theoretical Physics, Technikerstr. 21a, A-6020 Innsbruck, Austria}

\begin{abstract}
Collective behaviour in living systems is usually modelled as the outcome of a \emph{direct} social drive: agents are rewarded, or hard-wired, to align with or approach their neighbours. Here we show that aggregation can instead emerge from an \emph{indirect} objective. We let reinforcement learning foragers, initially performing a random walk, optimize their dynamics from a purely individual reward for finding replenishable targets, while perceiving only their conspecifics and never the targets themselves. As the visual range grows, the agents undergo a sharp crossover from an environment-tuned individual search to a scale-agnostic collective one, and this crossover coincides with the onset of spatial aggregation. Thus a collective phase arises as a by-product of optimal foraging, without any direct reward for grouping. A minimal analytical first-passage model reproduces the transition as a crossover between the two search strategies. Our results identify indirect, resource-driven reward as a generic route to emergent collective phenomena.
\end{abstract}

\maketitle

\section*{Introduction}
Collective motion is one of the most conspicuous forms of self-organization in nature, from bird flocks and fish schools to bacterial swarms and active colloids~\cite{vicsek1995novel, ballerini2008interaction, marchetti2013hydrodynamics}. The dominant theoretical picture, from the Vicsek model onward, explains such order through a \emph{direct} social interaction: each agent is driven to align its heading with, or move towards, its neighbours~\cite{vicsek1995novel, couzin2002collective}. When learning is introduced into these models, the same logic is inherited, and agents are typically rewarded explicitly for matching the velocity of their neighbours, so that flocking is built into the objective~\cite{durve2020learning}, or learning is used to steer and control the collective state directly~\cite{falk2021learning}. In all of these approaches, the collective state is what the agents are nudged to produce.

Here we propose a different route: can collective order emerge when nothing in the objective refers to it? Animals rarely receive a reward for grouping as such, but rather for finding food, mates or shelter, and social structure arises as a by-product of pursuing those individual goals~\cite{couzin2002collective, sims2008scaling}. Foraging is the paradigmatic example: a forager gains from locating resources, and the presence of successful conspecifics is at best an \emph{indirect} cue to where those resources are~\cite{torney2009context}. Whether optimizing such an indirect objective is enough to generate collective phenomena, and of what kind, is the question we address.

Finding the optimal strategy in a given environment is already a hard problem, amounting to a search over an infinite family of possible step-length distributions; introducing a population of interacting agents makes it harder still, since the optimum for each agent now depends on the strategies adopted by all others, and analytical approaches break down. We therefore turn to a reinforcement learning (RL) formulation of foraging, in which strategies are optimized through interaction with the environment rather than by assuming direct access to the objective function.
This approach has recently emerged as a versatile alternative to costly direct optimization~\cite{cai2025reinforcement, munoz2025learning, caraglio2024learning}, and has been applied across a broad range of problems, including collective settings~\cite{loffler2023collective, grauer2024optimizing}.
We build up from a single-agent framework, in which agents learn to adapt their dynamics from random walks to optimal search strategies~\cite{munoz2024optimal}, and extend it to a population of independent learners that share no information and interact only through vision. Each agent is rewarded solely for finding one of the replenishable targets present in the environment; crucially, it cannot see the targets, only whether nearby conspecifics have recently captured one or not.

\begin{figure}[b]
\centering
    \includegraphics[width=0.6\columnwidth]{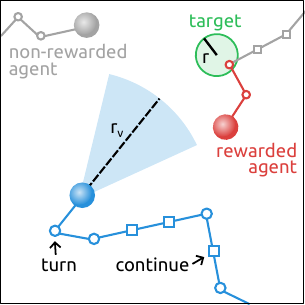}
    \caption{\textbf{Model.} An agent (blue) moves by \emph{continue} and \emph{turn} actions and perceives conspecifics within a vision cone of half-width $\pi/8$ and range $\VR$. Targets (green, radius $r$) are \emph{not} visible to the agents. A tagged agent (red) has just collected a target and is visible as rewarded for a time $\tau_R$; grey agents are non-rewarded.}
    \label{fig:scheme}
\end{figure}

We find that, as the visual range increases, the agents spontaneously switch from an individual, environment-tuned search to a collective one, and that this switch is accompanied by a transition from a disordered to an aggregated spatial state. The collective phase is therefore not designed in, but discovered by the agents as the optimal response to an indirect reward. More broadly, our results position reinforcement learning as a promising tool for uncovering emergent physics in regimes where direct optimization and analytical theory cannot reach.

\section*{Methods}%

We consider $N_a$ agents moving in a two-dimensional square box with side length $L = 50$ and periodic boundary conditions (see \cref{fig:scheme}). They start an episode at a random position and advance in steps of unit length. After each step, they either \emph{continue} in the current direction ($\uparrow$) or \emph{turn} ($\Rsh$) in a random direction. $N_t$ immobile targets of radius $r$ are placed uniformly at random. An agent collects a target when it comes within $r = 0.5$ of it, after which the target is depleted and unavailable for a time $\tau$. After a collection, the responsible agent is \emph{tagged} as rewarded for a time $\tau_R$. We contrast two scenarios: a \emph{competitive} one, in which a collected target is unavailable to \emph{all} agents during $\tau$, and a \emph{cooperative} one, in which the depletion applies only to the collecting agent, so the target remains available to others. 

The essential ingredient is the information available to the agents: they perceive other agents but never the targets. Vision is restricted to a cone of half-width $\pi/8$ and range $\VR$, and is coarse-grained into three states: no agent in view ($v=0$), at least one non-rewarded agent in view ($v=1$), or at least one rewarded agent in view ($v=2$). Cases where both rewarded and non-rewarded agents are in view get $v=2$.  Social cues are therefore the only channel through which any information about target locations can reach an agent.

\textbf{Searchers as RL agents }%
Each agent chooses its action from a policy $\pi(a|s)$, the probability of action $a$ given state $s=[c,v]$. The counter $c$ records the number of steps since the last turn; as shown in Ref.~\cite{munoz2024optimal}, conditioning on $c$ lets an agent reproduce a walk with \emph{any} step-length distribution $p(\ell)$, and in particular the L\'evy and bi-exponential walks that are near-optimal for foraging. The visual input $v\in\{0,1,2\}$ encodes the three cues above. We train $\pi(a|s)$ by RL, rewarding $R=1$ for each target collected, using Projective Simulation~\cite{briegel2012projective}, a model-free algorithm (see \cref{app:training}). Each of the $N_a$ agents carries and updates its own policy independently, and agents interact only through their visual cue. Training proceeds in episodes of $5000$ steps, with the target distribution randomly resampled at the start of each episode.

%
\section*{Results}
\begin{figure}
\centering
    \includegraphics[width=0.7\columnwidth]{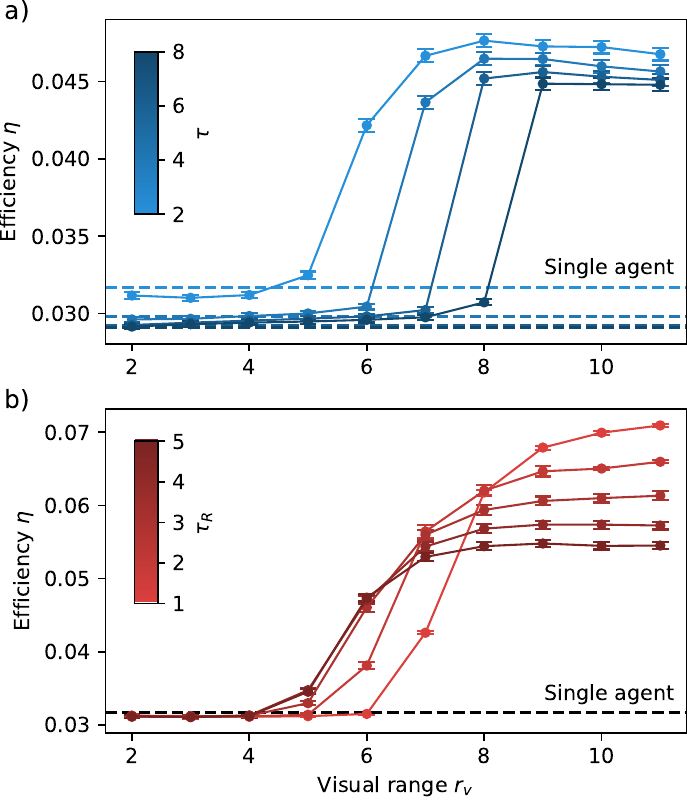}
    \caption{\textbf{Foraging efficiency.} Mean reward per step $\eta$ versus visual range $\VR$ in the cooperative regime. (a) Fixed tag time $\tau_R=9$, several depletion times $\tau$. (b) Fixed $\tau=2$, several tag times $\tau_R$. Below a $\tau$- and $\tau_R$-dependent visual range the agents recover the blind single-agent efficiency (dashed lines); above it, $\eta$ jumps and plateaus. Error bars: SEM over 4 runs of $N_a=50$ agents.}
    \label{fig:rewards}
\end{figure}

\textbf{An efficiency transition } We start by training $N_a=50$ in an environment with  $N_t=100$ targets, exploring different depletion times $\tau$, tag times $\tau_R$ and visual ranges $\VR$. Although the agents learn independently, they converge to nearly identical rewards and policies; we therefore report averages over the $N_a$ agents and, to assess training robustness, over 4 independent initializations
(see \cref{app:training}).

\cref{fig:rewards}a shows the learned efficiency $\eta$, the mean reward per step, as a function of $\VR$ for several $\tau$ for the cooperative scenario. A striking pattern emerges: at small $\VR$ every agent attains the efficiency of a single agent trained in the same environment \emph{without} vision, as if the social channel were useless. Small deviations are due to finite training times and the larger state space of the visual agents, which makes their training effectively harder. Then, at a well-defined $\VR$, $\eta$ jumps sharply and saturates at a higher plateau. The same transition appears when $\tau$ is fixed and $\tau_R$ is varied (\cref{fig:rewards}b): below a threshold $\VR$ all curves collapse onto the blind value, and above it $\eta$ rises. Larger $\tau_R$ shifts the jump to smaller $\VR$ and lower plateau. This reflects a trade-off in what a tag means: a short $\tau_R$ marks an agent that is still next to the target it just found, a reliable but rarely seen cue, whereas a long $\tau_R$ marks agents that may already have drifted far from the target, a frequent but noisier cue.

\textbf{Two dynamical strategies }%
\begin{figure}
    \includegraphics[width=\columnwidth]{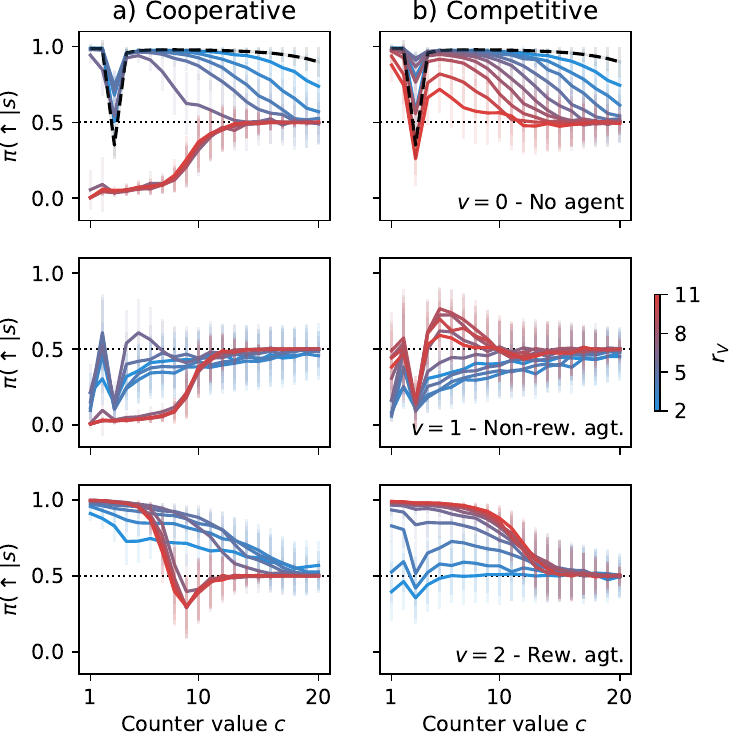}
    \caption{\textbf{Learned policies.} Continuing probability of the learned policies $\pi$ for the three visual states, in the competitive and cooperative scenarios, at $\tau=4$, $\tau_R=9$, $N_a=50$. Dashed lines in no agent state: single-agent (blind) policy. Dotted lines: initial policy. Error bars: standard deviation over 4 runs and $N_a$ agents (collective) and over $12000$ agents (single).}
    \label{fig:hmats}
\end{figure}
To understand the jump, we inspect the learned policies (\cref{fig:hmats}) for $\tau=4$ and $\tau_R=9$, resolved by depletion rule and visual state $v$. Consider first the \emph{no-agent} state $v=0$, compared to the single-agent strategy (dashed). As shown previously~\cite{munoz2024optimal}, the blind optimum has a high continuing probability with a marked drop at counter $c\sim\tau$, the signature of a scale-adaptive walk argued to be optimal here~\cite{ferreira2021landscape}. In the cooperative case at small $\VR$, and in the competitive case throughout, the agents adopt precisely this strategy, exploiting the length scales set by the environment. In the cooperative case at large $\VR$, however, the strategy inverts: the agents now turn with high probability at every counter value, performing an unbiased random walk. The same reversal appears in the \emph{non-rewarded} state, and, importantly, it sets in at the same $\VR=6$ where the efficiency jumps in \cref{fig:rewards}. The competitive scenario does not invert here, but does so for other environmental parameters, as we will show below.

The \emph{rewarded} state $v=2$ completes this picture. In the cooperative case it is always beneficial to follow a tagged agent, since a live target sits nearby; in the competitive case at small $\VR$ it confers no advantage, and the policy stays close to its initial value, because the target the tagged agent just took is depleted for all other agents. As $\VR$ grows past $\sim\tau$, the tagged agent one sees is far enough away that the target may have revived by the time we reach it, and following again becomes worthwhile.

The \emph{non-rewarded} state $v=1$ shows a more intricate behaviour, entangled with the other two. 
In the cooperative case at $\VR\geq6$, its policy follows the same logic as the no-agent state $v=0$, turning at every counter. For smaller $\VR$, and in the competitive case throughout, a distinct policy emerges.
At $c=1$ agents mostly turn. At $c=2$ the continuing probability $\pi(\uparrow|s)$ increases, most clearly for the low-$\VR$ curves: trajectories reaching this counter have continued from $c=1$ and therefore arrive predominantly from the no-agent or rewarded visual states, so the policy here inherits their higher $\pi(\uparrow|s)$. At $c=3$ a dip appears, the same scale-adaptive feature found in the no-agent state, where agents exploit the characteristic length set by the depletion time $\tau$. Beyond it, the competitive curves ramp back up, before relaxing to $0.5$: by the same argument, trajectories reaching these counters arrive mostly from the rewarded state, whose policy they mirror. The cooperative curves show the same tendency, ramping up more gradually at larger $c$.

Together these policies define two distinct strategies. The first follows the single-agent behavior, exploiting environmental scales and hence mimicking a blind strategy. The second, active at large $\VR$, is collective and scale-agnostic: turn until a rewarded conspecific appears, then dash ballistically towards it. The agents converge to one or the other based on the environmental parameters, both in the cooperative and the competitive scenarios, although for the latter this happens at larger $\VR$.

\textbf{Emergence of aggregation }%
\begin{figure}
    \includegraphics[width=\columnwidth]{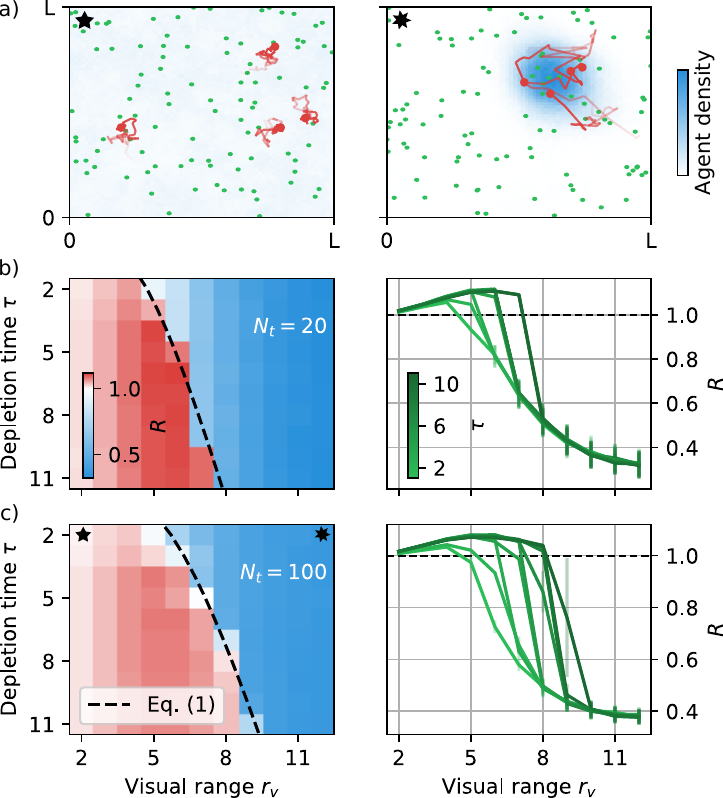}
    \caption{\textbf{Aggregation.} Results for an environment with $N_a = 25$, $\tau_R = 9$ and the cooperative scenario. a) Spatial distribution of agents over the final 5000 steps of a 15000 step run with $N_t = 100$ with trained policies at parameter values marked by the corresponding symbols in c. Four exemplary agent trajectories are overlaid. b, c) Clark--Evans index $R$ for two target densities $N_t$. Left panels: $R$ over the $(\tau,\VR)$ plane; dashed lines show \cref{eq:crossover} with $\ell=5,8$, respectively. Right panels: line cuts over previous plots; error bars: standard deviation over 4 runs and $N_a$ agents.}
    \label{fig:R_tau}
\end{figure}
The two strategies differ in how agents use social information, so we now ask whether the collective strategy produces genuine collective \emph{structure}. A first look at the agent density (\cref{fig:R_tau}a) already reveals that agents aggregate when following the collective strategy. We quantify spatial organization with the Clark--Evans aggregation index $R$, the ratio of the mean nearest-neighbour distance to that expected for a Poisson point process, so that $R<1$ signals clustering and $R>1$ overdispersion~\cite{clark1954distance}.

\cref{fig:R_tau} shows $R$ in the cooperative scenario at $\tau_R=9$ across the $(\tau,\VR)$ plane. The strategy transition seen in \cref{fig:rewards,fig:hmats} maps directly onto a transition in $R$: below it the population is spatially disordered ($R\approx1$), above it the agents aggregate, with $R$ decreasing as $\VR$ grows.
Remarkably, just \emph{before} the transition the system becomes over-dispersed, $R>1$. Indeed, as $\VR$ grows so does the chance of seeing another agent; in the dilute regime considered here these are mostly non-rewarded ($v=1$), implying a high turn probability  (\cref{fig:hmats}). This spaces agents out ever more strongly until the collective strategy arises and they cluster ($R<1$).

The transition is sharp and, for all $\tau$, the agents collapse onto a common $R$ both below and above the transition, confirming that only two macroscopic states exist, the individual and the collective, with no continuum of intermediate organizations. 

We note that this \emph{crossover} is not a thermodynamic phase transition: a finite-size analysis of a dense-phase order parameter, obtained by scaling the system at fixed density, locates a well-defined onset $\VR^{*}$ but shows that the width of the transition region does not close as the number of agents grows (see details in \cref{app:no_transition}). On the other hand, at higher target densities (i.e. larger $N_t$ in \cref{fig:R_tau}) the single-agent strategy remains competitive up to larger $\VR$: with targets easier to encounter even at large $\tau$, the blind searcher stays efficient, so the crossover to the collective strategy is pushed to higher $\VR$.

\begin{figure}
	\includegraphics[width=0.9\columnwidth]{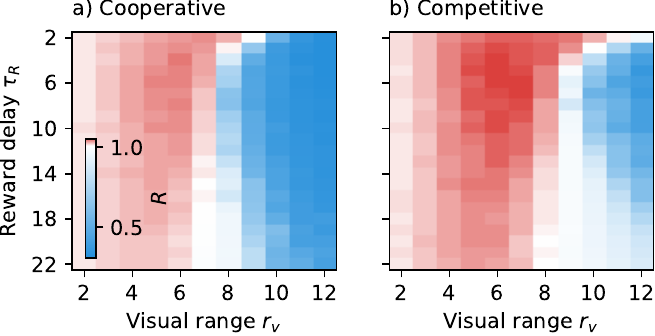}
	\caption{\textbf{Role of the depletion rule.} Clark--Evans index $R$ over the $(\tau_R,\VR)$ plane for the competitive and cooperative scenarios for $N_a=25$, $\tau=2$, and $N_t = 100$.}
	\label{fig:R_tauR}
\end{figure}

The two depletion rules make this mechanism explicit (\cref{fig:R_tauR}), here resolved as a function of the tag time $\tau_R$. Remarkably, the visual range at which the agents abandon the single-agent strategy remains almost constant across the entire $\tau_R$ window. The reason is that aggregation sets in as soon as heading towards a rewarded agent becomes more profitable than exploiting the scales of the environment, as the single-agent strategy does; this balance is set by the mean target spacing (here $d_t = 5$) and is largely insensitive to $\tau_R$. The threshold is higher in the competitive regime, where we further observe larger $R$ at intermediate $\VR$, signalling a stronger tendency of the agents to avoid one another. Finally, as $\tau_R$ grows the two strategies coexist over an increasingly wide range around the threshold, leaving the system disordered there ($R\approx1$).

\textbf{Physical modelling of the crossover }%
The transition can be rationalized as a crossover between the mean first-passage times (MFPTs) of the two identified strategies. To attain an analytical approach to the crossover, we consider an idealization of both strategies, considering the usual dilute assumption (see \cref{app:theory} for details). The single-agent case is modelled by the strategy considered optimal in this scenario~\cite{ferreira2021landscape}, an optimized bi-exponential walker. Its MFPT $T_s$ in two dimensions saturates at the ballistic mean free path $T_\infty = 1/(2r\rho)$ set by the target density $\rho$, with only a bounded $\mathcal{O}(r/\tau)$ correction~\cite{levernier2020inverse}. 

The collective strategy is modelled by a walker that diffuses with diffusion coefficient $D=1/4$ until it comes within a detection length $\lambda(\VR,\tau_R)$ of a target and performs a ballistic flight in its current direction. The latter models the event of finding a rewarded agent and performing the continue action. 
Solving the associated trapping problem gives an MFPT that grows only \emph{logarithmically} in the depletion time, $T_c \simeq (b^2/2D)\ln[\tau/(\lambda\ e^{-\ell/\lambda})]$. 

Equating this collective MFPT $T_c$ with the single-agent one $T_s$ yields the crossover detection radius $\lambda^{*}$, below which the individual strategy is faster and above which the collective one is,
\begin{equation}
	\lambda^*=  \frac{\ell}{W\left( \frac{\ell e^K}{\tau} \right )},
	\label{eq:crossover}
\end{equation}
where $W(\cdot)$ is the Lambert $W$ function, $K = \frac{\pi}{4r}\left(1-B\,\frac{r}{\tau}\right)$, and $B\approx0.64$ a numerical constant. Together, the detection radius $\lambda$ and the reactive length $\ell$ encode the compound process of observing a tagged agent and successfully capturing its target, hence encoding a dependence on $N_t$. In \cref{fig:R_tau} we test this prediction: taking $\lambda=\VR$ and $\ell=5,8$ for increasing $N_t$, the crossover of \cref{eq:crossover} qualitatively reproduces the boundary of the aggregation onset, showing that this minimal model captures the crossover. On the other hand, the competitive case (\cref{fig:R_tauR}) corresponds in turn to a larger $\ell$, since a target signalled by a rewarded agent is more likely to be already depleted, lowering the probability of capture upon reaching $\lambda$. Deriving closed forms for $\lambda$ and $\ell$, and hence for \cref{eq:crossover}, may be possible under certain assumptions and is left as an outlook of this work. We stress that the environments simulated here do not strictly satisfy the dilute condition under which the analytical MFPTs can be derived, and should therefore be read as identifying the mechanism that selects between the two strategies, and the scale at which it operates, rather than as a quantitative prediction of the boundary.

\section*{Discussion}
We have shown that a population of independent reinforcement learners, each optimizing only its own search efficiency and perceiving only its conspecifics, spontaneously develops a collective foraging strategy and spatial aggregation. The collective phase is not part of the \emph{direct} objective: no agent is rewarded for grouping, aligning, or approaching others. It emerges instead as the optimal response to an \emph{indirect} cue, the foraging success of neighbours, and appears abruptly, through a crossover in visual range that separates an environment-tuned individual search from a scale-agnostic collective one.

This shows that collective phenomena can emerge from an indirect source, in this case maximizing target acquisition. It stands in contrast to the standard route to collective motion, in which order is imposed through a direct social term, whether a hard-wired alignment rule or a reward for matching neighbours. Most importantly, that the same learning framework which reproduces optimal single-agent foraging~\cite{munoz2024optimal} also uncovers, unprompted, a genuine dynamical crossover suggests that RL can serve as a discovery tool for emergent physics in active systems, not merely as an optimizer.

Our framework also lets us probe how the form of the social interaction shapes this behaviour: depending on whether a depleted target is unavailable only to the agent that took it (cooperative) or to every agent (competitive), the searchers effectively share or compete for resources. We find that competition delays the onset of collective behaviour, pushing the crossover to the aggregated state to substantially larger visual ranges. More broadly, our results suggest that many instances of collective behaviour attributed to direct social drives may instead be by-products of individually optimal resource use, and that the character of the resource dynamics is what selects the collective state. 

A natural and particularly intriguing direction is inhomogeneous learning, where only a fraction of the population adapts while the rest follow fixed strategies. This raises questions such as how the learned policies adjust to the presence of agents with fixed behaviours, or what fraction of learning agents is required for a collective phenomenon such as aggregation to emerge. A further direction is to extend the framework to heterogeneous populations, mobile or spatially structured resources, and richer perceptual channels, clarifying how additional cues, closer to the sensory capabilities of real organisms, shape what is learned.

\bibliography{biblio}

\section*{Acknowledgments}
The authors acknowledge the use of large language models during this work: to assist in extending the existing single-agent simulation code of \cite{munoz2024optimal} to the multi-agent setting, to provide guidance in developing and analysing the theoretical model, including the finite-size characterisation that identifies the aggregation onset as a crossover, and to assist in drafting and editing the manuscript. All simulations, analyses, and conclusions were checked by the authors, who take full responsibility for the content.

This research was funded in part by the Austrian Science Fund (FWF) [SFB BeyondC F7102, DOI: 10.55776/F71; WIT9503323, DOI: 10.55776/WIT9503323]. For open access purposes, the authors have applied a CC BY public copyright license to any author accepted manuscript version arising from this submission. This work was also supported by the European Union (ERC Advanced Grant, QuantAI, No. 101055129; and ERC Consolidator Grant, MAPEI, No. 101001267). 
The views and opinions expressed in this article are however those of the author(s) only and do not necessarily reflect those of the European Union or the European Research Council - neither the European Union nor the granting authority can be held responsible for them.
GV acknowledges additional support from the Knut and Alice Wallenberg Foundation (grant number 2019.0079), and from the Göran Gustafsson Foundation for Research in Natural Sciences and Medicine.

\section*{Data availability}
All the code necessary to reproduce the results of this paper are available through the Python library \texttt{rl\_opts}~\cite{rlopts}.

\appendix
\input{appendix}

\end{document}

%% file: appendix.tex
\onecolumngrid

\section{Training}
\label{app:training}

Each agent is trained with Projective Simulation~\cite{briegel2012projective}, an off-policy reinforcement-learning algorithm in which the policy is stored as a weighted network of clip transitions. At every step the agent perceives its state $s=[c,v]$, samples an action from $\pi(a|s)$, and, upon collecting a target, reinforces the sequence of transitions that led to the reward. We refer to the appendix of \cite{munoz2024optimal} for a thorough introduction on the use of Projective Simulation for the foraging problem. For this work, for all shown results, the tunable parameters of the algorithm are set  to $\gamma = 10^{-5}$ and glow $\eta = 0.1$. Different parameters typically reach the same trained policies although at different training speeds. The chosen parameters robustly achieved good performances across all cases studied. The $N_a$ agents hold independent networks and never exchange weights, so all coupling is mediated by the shared environment through the visual state $v$. All policies reported here come from agents trained for $10^4$ episodes of $5000$ steps each, with the target field resampled at the start of every episode.

\section{Theoretical model for strategy transition}
\label{app:theory}
In this section, we develop a minimal model that explains the transition between the single (blind) and collective search strategies observed in the learning experiments. We note that our goal is not to develop an exact model that reproduces the behaviour of the simulated learning agents, but rather to showcase how the phase diagram for the optimal strategy can arise from the competition between a single and a collective strategy. For that, we make a set of assumptions that allow us to simplify the problem and obtain an analytical expression of the crossover $\VR^*$:
\begin{enumerate}
    \item The single-agent strategy is modelled by an optimized bi-exponential random walk. This is indeed the most accurate assumption, as prior work has demonstrated that such walk may be optimal for this type of search problem~\cite{ferreira2021landscape} and that learning agents can converge to these~\cite{munoz2024optimal}.
    \item The collective agent strategy is modelled by a walker with two dynamics: first, it performs a random walk with step length $d=1$. This mimics the situation in which the agent does not see any other agent and changes direction at every step. Then, the visualization of an agent is simplified by considering that, if the agent reaches a distance $x < \lambda$ from a target, it performs a ballistic walk in its current direction. $\lambda$ will be later related to the visual range $r_v$, as well as the success probability after observing a tagged agent. This models the situation in which the agents perform the continue action with very high probability after seeing a rewarded agent.
    \item We focus here on the dilute regime, i.e. $\rho = N_t / L^2 \rightarrow 0$. This is typically the hardest assumption to realize in the learning scenarios considered here: as the density decreases, the reward function becomes more sparse, effectively increasing the training difficulty. Nonetheless, our results show that even at $N_t = 20, L = 50$ and $\rho = 0.008$, an acceptable regime for the theory, the learning algorithm succeeds and hence its results can be compared to those presented  in this section.
    
\end{enumerate}

Our goal is to extract the mean first passage time (MFPT) of each strategy, and compute the advantage of one or the other as a function of the depletion time $\tau$ and visual range $\VR$.

\subsection{Setup}
We consider here a single walker looking for replenishable targets in a two dimensional box of side size $L\in \mathbb{R}$ with periodic boundary conditions. There are $N_t$ targets, giving a number density $\rho = N_t / L^2$. Each target is a disk of radius $r$. Lengths and times are measured in units of the step size $d = 1$ and of the speed, both set to one, so that a time interval equals the path length travelled in it.

When a target is acquired it is depleted for a time $\tau$, during which it cannot be reacquired. This phenomenon is analogous to the walker being displaced a distance $l_c$ after a target acquisition~\cite{levernier2020inverse}. In order to match the setup proposed in the main text, we will assume here that $\tau = l_c$. We note that this assumption will not have an effect on the general behaviour of the model presented below, and will only contribute as a negligible factor.

\subsection{Single agent search}

We start by considering a \emph{blind} agent that does not see other agents. In this case, its strategy is fully dependent on the properties of the environment, and in particular of $\tau$. In particular, and following recent work~\cite{ferreira2021landscape, munoz2024optimal}, we assume that the agent performs an optimized bi-exponential walk, performing a random walk with step lengths sampled from 
\begin{equation}
  p(\ell) = \sum_{i=1,2} \frac{\omega_i}{d_i}\, e^{-\ell/d_i},
  \qquad \omega_1 + \omega_2 = 1.
  \label{eq:biexp}
\end{equation}
It has been shown that this type of bi-exponential walk optimizes target acquisition whenever considering a short intensive scale $d_2$ and a long relocating scale $d_1 \gg d_2$. Although a bi-exponential step-length distribution has finite moments and is therefore asymptotically diffusive, with walk dimension $d_w = 2$, the asymptotic regime is not reached here. The relevant comparison is between the long scale $d_1$ and the mean free path
\begin{equation}
  T_{\infty} = \frac{1}{2 r \rho},
  \label{eq:mean_free_path}
\end{equation}
which is the mean path length travelled by a ballistic searcher before encountering a target of radius $r$ at density $\rho$. When $d_1 \gtrsim T_{\infty}$ the searcher typically encounters a target within a single flight and never turns before capture, so its motion is effectively ballistic on the scale of the search and hence its effective walk dimension is $d_w = 1$. This is the regime realized in our simulations.

\begin{figure}
\centering
    \includegraphics[width=\columnwidth]{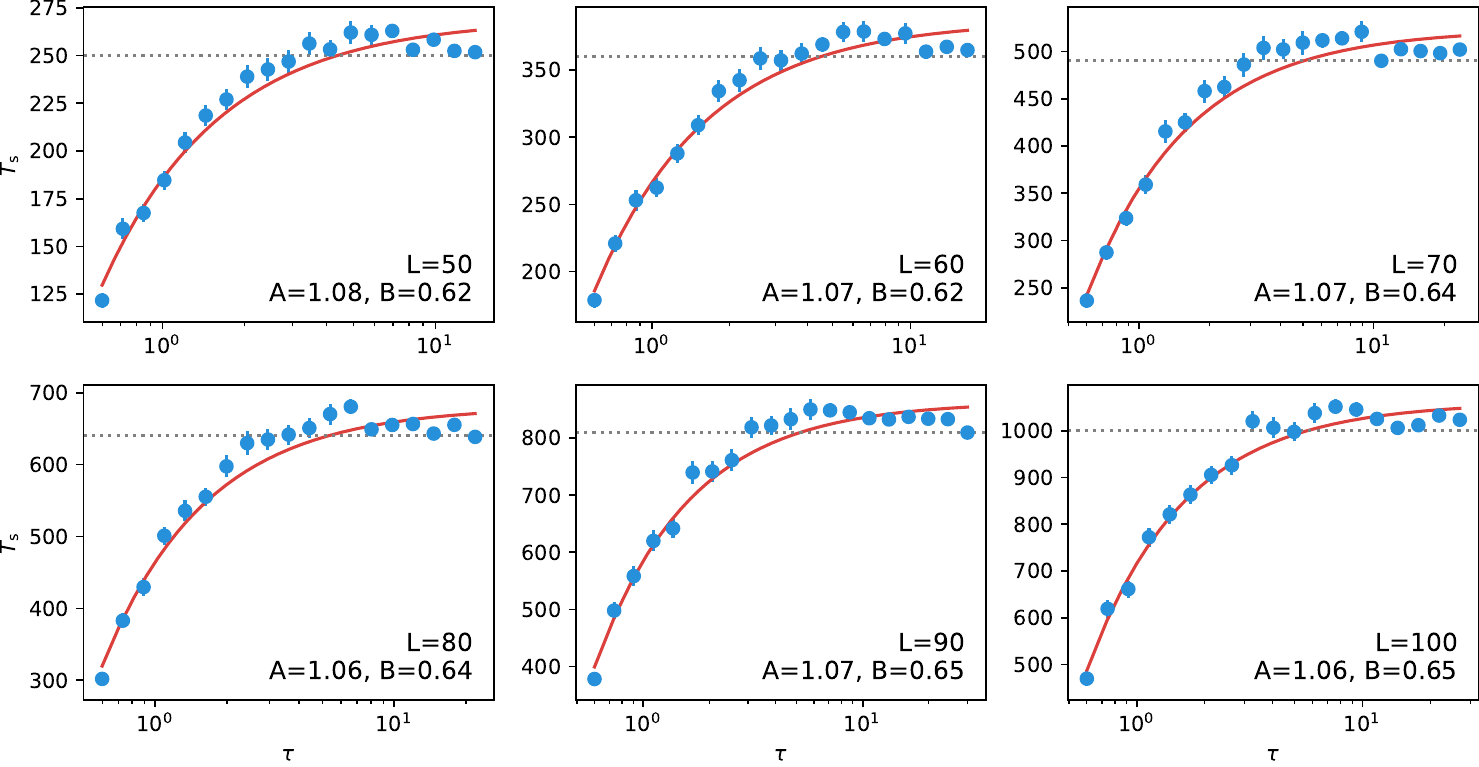}
    \caption{\textbf{Single agent MFPT} Scaling of the MFPT for the single agent as a function of the depletion time $\tau$, for $N_t = 10$, $r = 0.5$ and different box side lengths $L$. Blue data points show averages over $5\cdot 10^3$ simulations, with the errorbars showcasing the standard deviation of the mean. Solid lines are obtained from \cref{eq:mfpt_single} with fitted parameters $A,B$ shown in the plot. Dotted line shows the value of the  mean path length \cref{eq:mean_free_path}.}
    \label{app:fig_mfpt_single}
\end{figure}

Since we consider a spatial dimension $d_s = 2 > d_w = 1$, exploration is non-compact and the mean first-passage time of a walker starting at distance $x = \tau$ from the closest target is~\cite{levernier2017universal, levernier2020inverse}

\begin{equation}
  T_{\textrm{n-c}} \sim A\,\frac{V^{(d_w+\psi)/d_s}}{r^{\psi}}
  \left[1 - B\left(\frac{r}{\tau}\right)^{\psi}\right],
  \label{eq:mfpt_levernier}
\end{equation}
where $A,B$ are numerical constants, the volume $V=1/\rho$ and $\psi = D - d_w = 1$ is the transience exponent. Inserting these constants into the previous equation, we get

\begin{equation}
  T_s(\tau) = A T_{\infty}
  \left[1 - B\left(\frac{r}{\tau}\right)\right]. 
  \label{eq:mfpt_single}
\end{equation}
One can indeed see that the previous saturates to a value proportional to the mean free path $T_{\infty}$. Indeed, our numerical fittings of \cref{eq:mfpt_single} show both its validity and also that $A\sim 1$ and $B\approx 0.64$ in most cases. These results are shown in \cref{app:fig_mfpt_single} for different environmental parameters.

\subsection{Collective search strategy}
We discuss now the case of the collective search strategy. As commented above, we simplify this strategy by considering two different search phases:

\begin{enumerate}
    \item First, the walker performs an unbiased random walk with step length $d=1$ until reaching a distance $x < \lambda$ from the target. The previous is bounded by $\VR \leq \lambda \lesssim \VR+\sqrt{D\tau_R} +r$, where $D$ is the diffusion coefficient of the agents, and $\sqrt{D\tau_R}$ represents the mean displacement of an agent that captured the target in its furthest point (i.e. exactly at its boundary). In practice, as done in the main text, we would typically assume that $\lambda \approx \VR$, although the results of these section to not assume the later and give a strict results for the crossover as a function of $\lambda$ rather than $\VR$.
    \item Second, the agent performs ballistic motion in its current direction.
\end{enumerate}

As commented above, this mimics the behaviour of the collective search strategy: when no other agents are seen, the focal agent turns with probability $p_t \sim 1$. Then, after seeing a rewarded agent, it continues with probability $p_c \sim 1$. 

\subsubsection{Deterministic absorption}
We consider now an educational example, that will help us set the foundations for the collective search MFPT. Let us assume here that whenever the agent is at $x<\lambda$, it performs a ballistic flight that will encounter the target with probability 1. This means that the boundary at $\lambda$ is effectively absorbing. This assumes that the walker can reorientate towards the target direction, an assumption that will be dropped later. In this case, the MFPT is hence related to the unbiased random walk part of the walker's dynamic, as the dynamic part contributes with a fixed time $\lambda$.

The problem can then be rephrased as a two dimensional trapping problem, solved in the Wigner Seitz approximation: each target occupies on average an area $1/\rho$, which we replace by a disk (the cell) of radius
\begin{equation}
  b = \frac{1}{\sqrt{\pi \rho}} = \frac{L}{\sqrt{\pi N_t}} .
  \label{eq:cell}
\end{equation}
The searcher diffuses inside this cell, is absorbed at the detection circle
$x = \lambda$, and is reflected at the cell edge $x = b$ (i.e. the agent would enter an identical cell of the neighbour target, which is effectively the same as being reflected within the current cell). The MFPT $T(x)$ from radius $x$ obeys $D\nabla^2 T = -1$, where $D$ is a diffusion coefficient, which in radial form
reads
\begin{equation}
  \frac{D}{x}\frac{d}{dx}\!\left(x \frac{dT}{dx}\right) = -1,
  \qquad T(\lambda) = 0, \qquad T'(b) = 0 .
  \label{eq:radial}
\end{equation}
The diffusion constant follows from the microscopic walk: with unit steps and
unit speed, after $N$ steps $\langle R^2\rangle = N d^2$ while the elapsed time
is $t = N d$, so $\langle R^2\rangle = d\,t$. Comparing with the two dimensional diffusion
law $\langle R^2\rangle = 4Dt$ gives
\begin{equation}
  D = \frac{d}{4} = \frac14 .
  \label{eq:D}
\end{equation}
Integrating twice with the two boundary conditions we obtain
\begin{equation}
  T(x) = \frac{\lambda^2 - x^2}{4D} + \frac{b^2}{2D}\,\ln\frac{x}{\lambda} .
  \label{eq:Tannulus}
\end{equation}
Evaluated at the start distance $x = \tau$, and inserting Eq.~\eqref{eq:D} and
$b^2 = L^2/(\pi N_t)$,
\begin{equation}
  T_{\mathrm{det}}(\tau)
  = \bigl(\lambda^2 - \tau^2\bigr) + \frac{2 L^2}{\pi N_t}\,\ln\frac{\tau}{\lambda}.  \label{eq:T_deterministic}
\end{equation}
In the dilute regime $\lambda, \tau \ll b$ the logarithm dominates and
$T_{\mathrm{det}} \simeq A \ln(\tau/\lambda)$ with the prefactor
\begin{equation}
  A \equiv \frac{2 L^2}{\pi N_t} = \frac{2}{\pi \rho} = 2 b^2 .
\end{equation}
The previous result is indeed consistent with known results on first passage times in confined scale invariant system~\cite{condamin2007first}: in $D = 2$ the
MFPT grows linearly with the confining area $1/\rho$ and logarithmically with
the source to target separation. 

\subsubsection{Probabilistic absorption}
We now relax the assumption of deterministic absorption at $x = \lambda$. In our original problem, there are two main sources that may prevent agents in this second phase from reaching the target. First, the current direction of the agent may make it miss the target, even at $x\leq\lambda$. Indeed, the later probability, when starting the ballistic flight at $x = \lambda$ can be found from pure geometric considerations as $p_c=\min\!\bigl(\theta,\; 2\arcsin(r/\lambda)\bigr)/(2\pi)$, where $\theta$ is the width of the visual cone (set for all the numerical results in the main text to $\theta = \pi/4$). Second, the agent does not see an immobile target, but rather agents that have been tagged after acquiring the target. In principle, in the dilute agent regime, one can consider that a single agent exists around each target, so that after acquisition, agents are in the first random walk dynamical phase. This means that encountering a tagged agent relates to a signal that diffuses from the target with diffusion coefficient $D$.

Due to the complexity of the previous consideration, we leave the derivation of the exact MFPT for such a case for future work. Instead, we consider a simplification, sufficient to qualitatively recover the numerical results presented in the main text: we consider that the boundary at $\lambda$ is \emph{partially} absorbing. This means that, only a fraction of the agents reaching the boundary will actually end up reaching the target. In the previous framework, this entails considering a Robin boundary condition at $T(\lambda)$:
\begin{equation}
  \left.\frac{dT}{dx}\right|_{x=\lambda} = \frac{1}{\ell}\,T(\lambda),
  \label{eq:robin}
\end{equation}
where $\ell$ is a reactive length: $\ell \to 0$ recovers perfect absorption while
$\ell \to \infty$ is a perfectly reflecting, never capturing, boundary. Physically
$\ell$ bundles the effects mentioned above. Comparing to our learning framework, larger $\ell$ can be for instance related to larger tag times $\tau_R$ or changes in the agent's density.

Solving the \cref{eq:radial} by considering the boundary conditions \cref{eq:robin} at $x = \lambda$, and keeping the leading order in the dilute limit
$b \gg \lambda$, we obtain the MFPT for the collective strategy $T_c$,
\begin{equation}
  T_{c}(\lambda, \tau) \simeq \frac{b^2}{2D}
  \left[\ln\frac{\tau}{\lambda} + \frac{\ell}{\lambda}\right]
  = \frac{b^2}{2D}\,\ln\frac{\tau}{\lambda\ e^{-\ell/\lambda}} .
  \label{eq:mfpt_collec}
\end{equation}
As we can see, the reactive length enters exactly as a rescaling of the ballistic radius $\lambda$. Indeed, we can define an effective radius $\lambda_{\mathrm{eff}} = \lambda\, e^{-\ell/\lambda}$ and map the directional problem onto an isotropic, perfectly absorbing problem with a wider trapping radius.

\begin{figure}
	\centering
	\includegraphics[width=0.5\columnwidth]{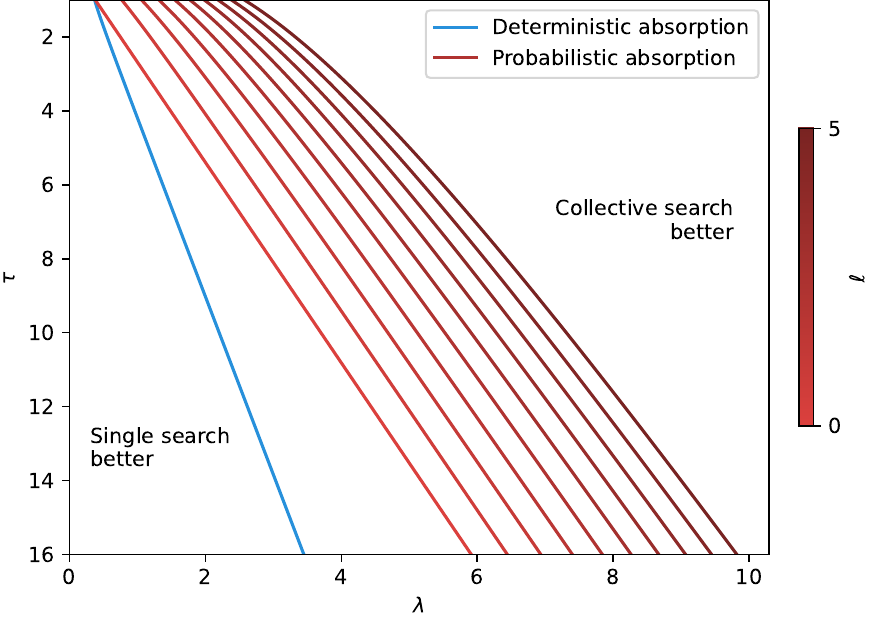}
	\caption{\textbf{Single-Collective strategy crossover} Blue line shows the crossover between the single and collective strategy when the former follows \cref{eq:T_deterministic}. The red lines show \cref{eq:crossover_app} for different values of $\ell$ and $B=0.64$.}
	\label{app:fig_crossover}
\end{figure}

\subsection{Dynamics phase diagram}
We now compare the MFPT obtain for the single and collective search strategies, namely \cref{eq:mfpt_single} and \cref{eq:mfpt_collec}, respectively. Our aim is to recover the phase diagram found by the learning agents in \cref{fig:R_tau}. For that, we define $\alpha = T_c(\lambda, \tau) / T_s(\lambda)$. Then, equating the previous to one, we find that the transition between two strategies, i.e. the value of $\lambda$ at which the two MFPTs are equal,
\begin{equation}
    \lambda^*=  \frac{\ell}{W\left( \frac{\ell e^K}{\tau} \right )},
    \label{eq:crossover_app}
\end{equation}
where $W(\cdot)$ is the Lambert W function, and $K = \frac{\pi}{4r}\left(1-B\frac{r}{\tau} \right)$. For $\lambda < \lambda^*$, the single strategy dominates as it has a lower MFPT, while the opposite happens at $\lambda > \lambda^*$.  This boundary is shown in \cref{app:fig_crossover} for different values of $\ell$, as well as the crossover boundary in the case in which $\alpha$ is computed with $T_{\mathrm{det}}(\tau)$ (\cref{eq:T_deterministic}) instead of $T_c(\lambda,\tau)$. As shown, increasing $\ell$ moves the boundary towards larger $\lambda$. Indeed, larger values of $\ell$ correspond to lower probabilities of actually finding the target when starting the ballistic flight. This makes the single search dominate over the collective one for larger $\VR$.

\section{Sharp crossover and dynamical coexistence}
\label{app:no_transition}
\begin{figure*}
	\centering
	\includegraphics[width=\textwidth]{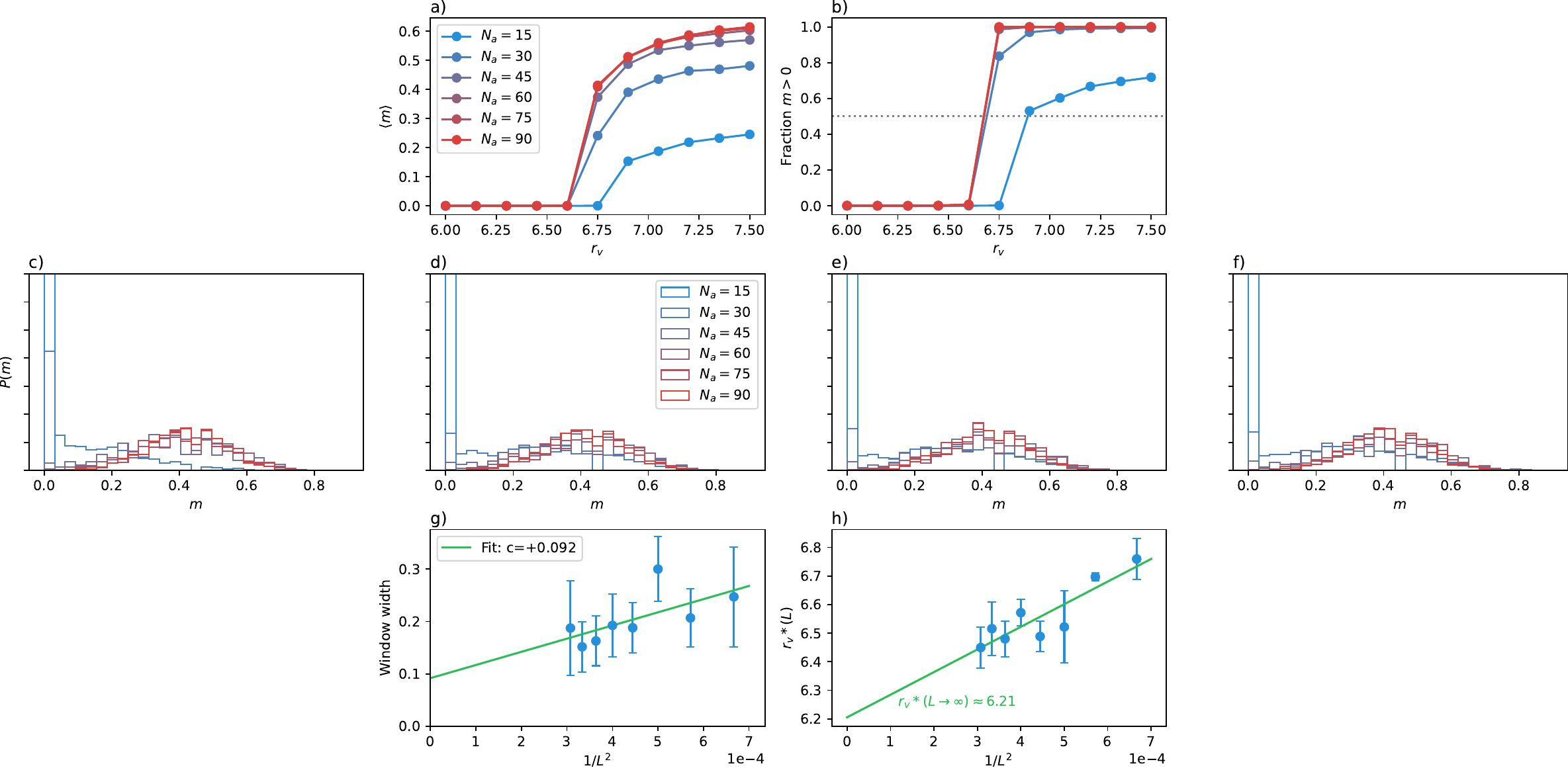}
	\caption{\textbf{Finite-size analysis of the aggregation onset.}
		Dense-phase order parameter $m$ measured over 8000 random initializations of policies for four independent trainings per $N_a$, with the system scaled at fixed agent and target density ($L\propto\sqrt{N_a}$).
		\textbf{(a)}~Mean order parameter $\langle m\rangle$ and \textbf{(b)}~fraction of runs reaching the aggregated state ($m>0$) versus visual range $r_v$.
		\textbf{(c)--(f)}~Distribution $P(m)$ near $r_v^{*}$ ($r_v = 7$) for four independently trained policies per parameter set.
		\textbf{(g)}~Width of the transition region (the $r_v$ interval over which the occupied fraction rises from $0.1$ to $0.9$) versus $1/L^2$; the fit extrapolates to a finite intercept $c\approx0.09$.
		\textbf{(h)}~Onset location $r_v^{*}(L)$ versus $1/L^2$, converging to $r_v^{*}(L\to\infty)\approx 6.2$.
		Error bars: spread across the four policies.}
	\label{app:fig_no_transition}
\end{figure*}

To characterize the onset of aggregation we define a dense phase order
parameter $m$, the fraction of agents with at least $N_\ell = 5$ neighbours
within a fixed radius $r_\ell = 2.5$. This threshold lies far above the
Poisson expectation $\rho_a \pi r_\ell^2 \approx 0.4$, so a
spatially random population has vanishing $m$.
We evaluate it over 8000 random initializations of four sets of $N_a$ trained policies, scaling the system at fixed agent and target densities ($\rho_a$, $\rho_t$ constant, $L\propto\sqrt{N_a}$) so that only the size grows. Both the mean order parameter $\langle m\rangle$ (\cref{app:fig_no_transition}(a)) and the fraction of runs that reach the aggregated state (\cref{app:fig_no_transition}(b)) rise sharply at a well-defined onset $\VR^{*}$, and the rise steepens with system size.

Near $\VR^{*}$ the distribution of $m$ is bimodal (\cref{app:fig_no_transition}(c)--(f)): a peak at $m=0$ (dispersed) coexists with a peak at finite $m$ (aggregated), so a single frozen policy relaxes into either state depending on its initial condition. This coexistence, and its weight, is reproduced across the four independently trained policies shown in panels (c)--(f), demonstrating that it is a property of the collective dynamics rather than of the learning. Larger systems fall into the aggregated state more reliably, which is what sharpens the onset.

Whether this onset is a genuine phase transition is decided by the width of the transition region, i.e. the interval of $\VR$ over which the fraction of occupied runs, over different environment initializations, rises from $0.1$ to $0.9$. A finite-size analysis (\cref{app:fig_no_transition}(g)) shows this width saturating at a finite value $c\approx0.09$ as $1/L^2\to0$ rather than approaching zero, identifying the onset as a sharp crossover rather than a thermodynamic phase transition. The onset location itself (\cref{app:fig_no_transition}(h)) converges to $\VR^{*}(L\to\infty)\approx6.2$.

%% file: biblio.bib
@article{munoz2024optimal,
  title={Optimal foraging strategies can be learned},
  author={Mu{\~n}oz-Gil, Gorka and L{\'o}pez-Incera, Andrea and Fiderer, Lukas J and Briegel, Hans J},
  journal={New Journal of Physics},
  volume={26},
  number={1},
  pages={013010},
  year={2024},
  publisher={IOP Publishing}
}

@article{levernier2020inverse,
  title={Inverse square {L{\'e}vy} walks are not optimal search strategies for d $\geq$ 2},
  author={Levernier, Nicolas and Textor, Johannes and B{\'e}nichou, Olivier and Voituriez, Rapha{\"e}l},
  journal={Physical Review Letters},
  volume={124},
  number={8},
  pages={080601},
  year={2020},
  publisher={APS}
}

@article{briegel2012projective,
  title={Projective simulation for artificial intelligence},
  author={Briegel, Hans J and De las Cuevas, Gemma},
  journal={Scientific reports},
  volume={2},
  number={1},
  pages={1--16},
  year={2012},
  publisher={Nature Publishing Group}
}

@article{ferreira2021landscape,
  title={Landscape-scaled strategies can outperform {L{\'e}vy} random searches},
  author={Ferreira, J and Raposo, EP and Ara{\'u}jo, HA and da Luz, MGE and Viswanathan, GM and Bartumeus, Frederic and Campos, Daniel},
  journal={Physical Review E},
  volume={103},
  number={2},
  pages={022105},
  year={2021},
  publisher={APS}
}

@article{sims2008scaling,
  title={Scaling laws of marine predator search behaviour},
  author={Sims, David W and Southall, Emily J and Humphries, Nicolas E and Hays, Graeme C and Bradshaw, Corey JA and Pitchford, Jonathan W and James, Alex and Ahmed, Mohammed Z and Brierley, Andrew S and Hindell, Mark A and others},
  journal={Nature},
  volume={451},
  number={7182},
  pages={1098--1102},
  year={2008},
  publisher={Nature Publishing Group}
}

@article{levernier2017universal,
	title = {Universal first-passage statistics in aging media},
	author = {Levernier, N. and B\'enichou, O. and Gu\'erin, T. and Voituriez, R.},
	journal = {Phys. Rev. E},
	volume = {98},
	issue = {2},
	pages = {022125},
	numpages = {20},
	year = {2018},
	month = {Aug},
	publisher = {American Physical Society},
	doi = {10.1103/PhysRevE.98.022125},
	url = {https://link.aps.org/doi/10.1103/PhysRevE.98.022125}
}

@article{condamin2007first,
  title={First-passage times in complex scale-invariant media},
  author={Condamin, SBENICHOU and B{\'e}nichou, O and Tejedor, V and Voituriez, R and Klafter, Joseph},
  journal={Nature},
  volume={450},
  number={7166},
  pages={77--80},
  year={2007},
  publisher={Nature Publishing Group UK London}
}

@article{vicsek1995novel,
  title={Novel type of phase transition in a system of self-driven particles},
  author={Vicsek, Tam{\'a}s and Czir{\'o}k, Andr{\'a}s and Ben-Jacob, Eshel and Cohen, Inon and Shochet, Ofer},
  journal={Physical Review Letters},
  volume={75},
  number={6},
  pages={1226--1229},
  year={1995},
  publisher={American Physical Society}
}

@article{couzin2002collective,
  title={Collective memory and spatial sorting in animal groups},
  author={Couzin, Iain D and Krause, Jens and James, Richard and Ruxton, Graeme D and Franks, Nigel R},
  journal={Journal of Theoretical Biology},
  volume={218},
  number={1},
  pages={1--11},
  year={2002},
  publisher={Elsevier}
}

@article{ballerini2008interaction,
  title={Interaction ruling animal collective behavior depends on topological rather than metric distance: Evidence from a field study},
  author={Ballerini, Michele and Cabibbo, Nicola and Candelier, Raphael and Cavagna, Andrea and Cisbani, Evaristo and Giardina, Irene and Lecomte, Vivien and Orlandi, Alberto and Parisi, Giorgio and Procaccini, Andrea and others},
  journal={Proceedings of the National Academy of Sciences},
  volume={105},
  number={4},
  pages={1232--1237},
  year={2008},
  publisher={National Academy of Sciences}
}

@article{marchetti2013hydrodynamics,
  title={Hydrodynamics of soft active matter},
  author={Marchetti, M Cristina and Joanny, Jean-Fran{\c{c}}ois and Ramaswamy, Sriram and Liverpool, Tanniemola B and Prost, Jacques and Rao, Madan and Simha, R Aditi},
  journal={Reviews of Modern Physics},
  volume={85},
  number={3},
  pages={1143--1189},
  year={2013},
  publisher={American Physical Society}
}

@article{durve2020learning,
  title={Learning to flock through reinforcement},
  author={Durve, Mihir and Peruani, Fernando and Celani, Antonio},
  journal={Physical Review E},
  volume={102},
  number={1},
  pages={012601},
  year={2020},
  publisher={American Physical Society}
}

@article{falk2021learning,
  title={Learning to control active matter},
  author={Falk, Martin J and Alizadehyazdi, Vahid and Jaeger, Heinrich and Murugan, Arvind},
  journal={Physical Review Research},
  volume={3},
  number={3},
  pages={033291},
  year={2021},
  publisher={American Physical Society}
}

@article{torney2009context,
  title={Context-dependent interaction leads to emergent search behavior in social aggregates},
  author={Torney, Colin and Neufeld, Zoltan and Couzin, Iain D},
  journal={Proceedings of the National Academy of Sciences},
  volume={106},
  number={52},
  pages={22055--22060},
  year={2009},
  publisher={National Academy of Sciences}
}

@article{loffler2023collective,
  title={Collective foraging of active particles trained by reinforcement learning},
  author={L{\"o}ffler, Robert C and Panizon, Emanuele and Bechinger, Clemens},
  journal={Scientific Reports},
  volume={13},
  number={1},
  pages={17055},
  year={2023},
  publisher={Nature Publishing Group UK London}
}

@article{munoz2025learning,
  title={Learning to reset in target search problems},
  author={Mu{\~n}oz-Gil, Gorka and Briegel, Hans J and Caraglio, Michele},
  journal={New Journal of Physics},
  volume={27},
  number={9},
  pages={093701},
  year={2025},
  publisher={IOP Publishing}
}

@article{caraglio2024learning,
  title={Learning how to find targets in the micro-world: the case of intermittent active Brownian particles},
  author={Caraglio, Michele and Kaur, Harpreet and Fiderer, Lukas J and L{\'o}pez-Incera, Andrea and Briegel, Hans J and Franosch, Thomas and Mu{\~n}oz-Gil, Gorka},
  journal={Soft Matter},
  volume={20},
  number={9},
  pages={2008--2016},
  year={2024},
  publisher={The Royal Society of Chemistry}
}

@article{grauer2024optimizing,
  title={Optimizing collective behavior of communicating active particles with machine learning},
  author={Grauer, Jens and Jan Schwarzendahl, Fabian and L{\"o}wen, Hartmut and Liebchen, Benno},
  journal={Machine Learning: Science and Technology},
  volume={5},
  number={1},
  pages={015014},
  year={2024},
  publisher={IOP Publishing}
}

@article{cai2025reinforcement,
  title={Reinforcement learning for active matter},
  author={Cai, Wenjie and Wang, Gongyi and Zhang, Yu and Qu, Xiang and Huang, Zihan},
  journal={Biophysics Reviews},
  volume={6},
  number={3},
  year={2025},
  publisher={AIP Publishing}
}

@software{rlopts,
  author       = {Mu\~noz-Gil, Gorka and L\'opez-Incera, Andrea},
  title        = {\uppercase{RL}-\uppercase{O}pt\uppercase{S}: Reinforcement Learning of Optimal Search Strategies},
  month        = jan,
  year         = 2024,
  publisher    = {Zenodo},
  version      = {v1.0},
  doi          = {10.5281/zenodo.10450489},
  url          = {https://doi.org/10.5281/zenodo.7727873}}

@article{clark1954distance,
  title={Distance to nearest neighbor as a measure of spatial relationships in populations},
  author={Clark, Philip J and Evans, Francis C},
  journal={Ecology},
  volume={35},
  number={4},
  pages={445--453},
  year={1954},
  publisher={JSTOR}
}
